# Cluster of Dipolar Coupled Spins as a Quantum Memory Storage

A. K. Khitrin[1], V. L. Ermakov[2], and B. M. Fung[1]

[1]Department of Chemistry and Biochemistry, University of Oklahoma, Norman, Oklahoma 73019-0370, USA;

[2]Kazan Physical-Technical Institue, Kazan 420029, Russia.

*Abstract.* Spin dynamics of a cluster of coupled spins ½ can be manipulated to store and process a large amount of information. A new type of dynamic response makes it possible to excite coherent long-living signals, which can be used for exchanging information with a mesoscopic quantum system. An experimental demonstration is given for a system of 19 proton spins of a liquid crystal molecule.

Quantum information processing (QIP) is a fast developing field. Its various aspects are described in a recent monograph [1]. The idea of quantum computing, as it was originally proposed by Feynman [2], is the use of reversible quantum dynamics, instead of relaxational classical dynamics, for computing and simulations. The potential power of quantum systems for computing is based on their complicated dynamics. This complexity stems from the high dimensionality of the Hilbert space needed to represent the motion of even a small composite quantum system. The subsequent discovery of quantum algorithms, especially Shor's factorization algorithm [3], demonstrated that quantum computers can provide performance at a much higher level than classical computers. This has stimulated numerous attempts to build practical devices.

At present, nuclear magnetic resonance (NMR) is the most advanced technique for experimenting with QIP. The possibility of using bulk NMR for ensemble quantum computing is based on the idea of pseudopure states [4,5]. The most recent achievement is the implementation of Shor's algorithm to factorize the number 15 using a seven qubit (seven spins ½) system [6].

Despite obvious progress in the field, the prospects of building a large-scale quantum "supercomputer" are still unclear. The main problems in creating a universal quantum computer with many qubits are quantum decoherence and the necessity to individually address each single qubit. On the other hand, a less ambitious goal could be the use of quantum systems as elementary switches or co-processors within a computer, which is otherwise classical.

In order to realize the potential computational power, quantum co-processors should be large enough to have complicated dynamics but small enough to remain quantum. The number of qubits itself does not determine the complexity of a system. An "architecture" or the way these qubits are coupled is also important. As an example, ten uncoupled spins ½ is a trivial system. Ten spins with truncated J-couplings (the interactions used now in liquid-state NMR quantum computing) is a system with rather simple dynamics. However, simulating the dynamics of ten spins, coupled by dipole-dipole interactions, is near the limit of the performance of classical computers.

For a cluster of $N$ dipolar coupled spins, the maximum number of transitions $C_{2N}^{N+1} \sim 2^{2N}$ grows very fast with the increase in the number of spins. Even for a relatively small cluster of dipolar coupled nuclear spins a conventional NMR spectrum can be unresolved. Therefore, state-selective or qubit-selective pulses cannot be used and an alternative approach is needed to use spin clusters as quantum processors.

We have found that long and weak radiofrequency (RF) pulses can generate coherent long-living response signals in clusters of dipolar-coupled spins with unresolved spectra. This phenomenon makes it possible for the experimenter to "communicate" with a quantum system and to record a large amount of information in its spin state.

The physical system we used in our experiments is the nematic liquid crystal 4-*n*-pentyl-4'-cyanobiphenyl (5CB). Fast molecular motions average intermolecular dipole-dipole interactions between proton nuclear spins, but molecular orientational order leads to residual intramolecular interactions. Therefore, this system is a good example of an ensemble of identical non-interacting spin clusters. Inside each cluster, the 19 proton spins of the molecule are coupled with dipole-dipole interactions. A conventional $^1$H spectrum of 5CB obtained by the application of a "hard" 90° pulse has a width of about 25 kHz and does not display resolved peaks. However, we have found that a single sharp peak with a large amplitude can be produced by applying a very weak and long RF pulse. The width of this peak is three orders of magnitude less than that of the conventional spectrum. The frequency of the peak is exactly the same as that of the excitation pulse.

As a consequence of the huge number of states in our system ($2^{19}$) and an even larger number of allowed transitions, it is possible to simultaneously excite many coherent signals with a multi-frequency excitation. The spectrum in Fig. 1a shows the result of the simultaneous excitation of 110 peaks using a pulse which is the sum of the same number of circularly polarized harmonics. The frequency interval between neighboring harmonics is 200 Hz. The width of most of the peaks is about 12 Hz, but increases to about 20 Hz in the outer parts of the spectrum. It should be noted that the envelope of the peaks is almost a mirror image of the conventional spectrum.

Each of the peaks in the spectrum in Fig. 1a is phase-sensitive, so that any peak can be inverted to become positive by changing a sign of the amplitude of corresponding harmonic in the multi-frequency pulse. This makes it possible to encode a considerable amount of information in the resulting spectrum.

If we designate a negative peak as 0 and a positive peak as 1, the 110 peaks form a system of 110 bits. When they are used to store a number, it can take any value between 0 and $2^{110}-1$; i.e. a number with 33 decimal digits can be recorded. An alternative is to use every 5 bits to store a character of the English alphabet. Then, the 110 bits can be used to encode a phrase containing 22 characters. An example is illustrated by the spectrum in Fig. 1b.

Since the interactions between different molecules are averaged by fast molecular motions and we used only spatially homogeneous fields, the spin density matrix is a product of identical density matrices for individual molecules at all times. Therefore, the information is recorded in a single molecule and other molecules contain identical copies of the same information.

The idea of using spin echoes to store information in inhomogeneously broadened spectra was proposed by Erwin Hahn a long time ago [7]. However, without spin-spin interactions, the number of peaks (and the number of stored bits) can not exceed the number of spins in the system. In the presence of spin-spin interactions, the technique of spin echoes does not allow the manipulation of individual peaks.

Experiments on using such spin processor for parallel bitwise operations with long bit arrays are in progress and the results will be published elsewhere.

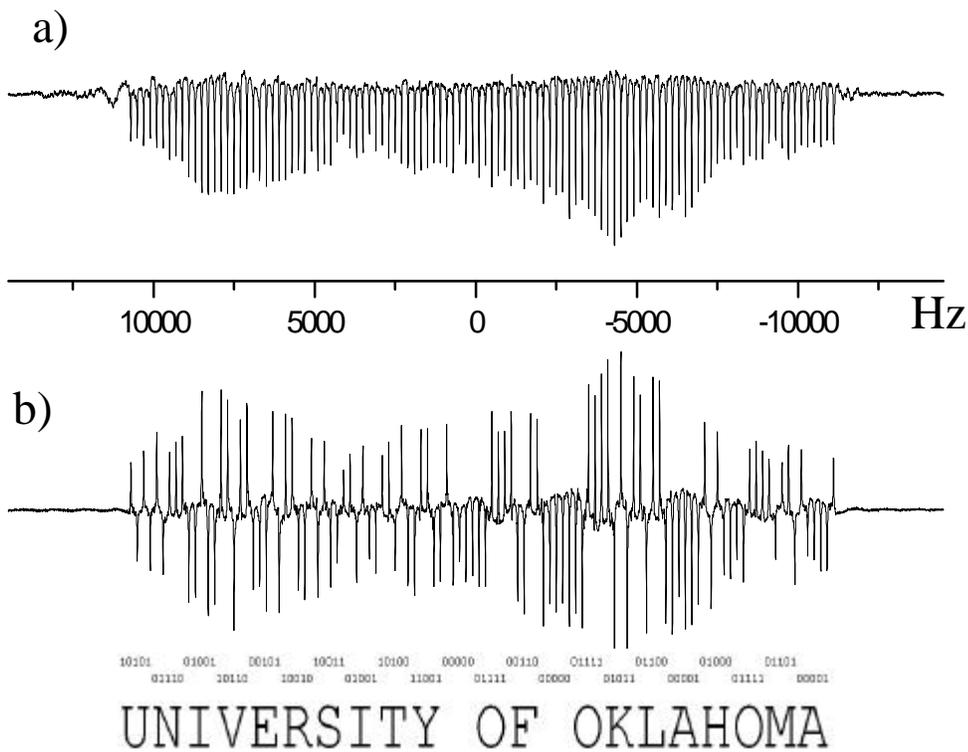

Fig. 1. $^1$H NMR spectra of 5CB obtained with multi-frequency pulses containing 110 harmonics. The RF amplitude of each of the harmonics was 3 Hz. The number of transients was 512; an acquisition delay of 1 ms was used to reduce the broad components. (a) The phases of all the harmonics are the same. (b) The amplitudes of some of the harmonics are inverted for the purpose of information storage, so that the peak is negative when a bit takes the value of 0 and positive when it takes the value of 1. Each character is coded with 5 bits by using successive binary numbers ("space"=00000, a=00001, b=00010, *etc.*); the binary codes and corresponding letters are displayed under the spectrum.